\documentclass{IEEEtran}

\usepackage{amsmath,amssymb,amsfonts}
\usepackage{algorithmic}
\usepackage{graphicx}
\usepackage{textcomp}
\usepackage{romannum}
\usepackage{amsmath}
\usepackage{csvsimple}
\usepackage{float}
\usepackage{xcolor}
\usepackage{url}
\usepackage{makecell}
\usepackage{subcaption}
\usepackage[export]{adjustbox}

\def\BibTeX{{\rm B\kern-.05em{\sc i\kern-.025em b}\kern-.08em
		T\kern-.1667em\lower.7ex\hbox{E}\kern-.125emX}}
\begin{document}
	\title{Propagation Channel Modeling by Deep learning Techniques}
	\author{Shirin Seyedsalehi, Vahid Pourahmadi, Hamid Sheikhzadeh, Ali Hossein Gharari Foumani}
	
	\maketitle
	
	\begin{abstract}
		Channel, as the medium for the propagation of the electromagnetic waves, is one of the most important parts of a communication system. Being aware of how the channel affects the propagation waves is essential for designing, optimization and performance analysis of a communication system. For this purpose, a proper channel model is needed. This paper presents a novel propagation channel model which considers the time-frequency response of the channel as an image. It models the distribution of these channel images using deep convolutional generative adversarial networks. Moreover, for the measurements with different user speeds, the user speed is considered as an auxiliary parameter for the model. StarGAN as an image-to-image translation technique is used to change the generated channel images with respect to the desired user speed. The performance of the proposed model is evaluated using existing metrics. Furthermore, to capture 2D similarity in both time and frequency, a new metric is introduced. Using this metric, the generated channels show significant statistical similarity to the measurement data.
	\end{abstract}
	

	\section{Introduction}
\label{sec:introduction}
{Propagation} channel modeling is one of the most essential parts of communication system design and simulation. When channel state information is available, transfer characteristics like signal constellation and allocated power can be adjusted in order to increase the system performance. A proper propagation model is needed for performance analysis of the system as well. 

Several methods have been developed for propagation modeling, which can be divided into deterministic and stochastic approaches \cite{wang2018survey}.
The first class of approaches tends to reconstruct the propagation process by solving Maxwell equations in a particular medium with given boundary conditions. Ray Tracing (RT) based models are the most commonly employed models of this type \cite{degli2009speed}. Deterministic approaches are accurate, but they need complex calculations and are heavily dependent on the exact description of the medium.
On the other hand, stochastic approaches try to provide a statistical description of channel characteristics, e.g., power delay profile (PDP). They can themselves be divided into two categories: Geometry-Based Stochastic Models (GBSMs) and Non-Geometry Based Stochastic Models (NGSMs). 

GBSM approaches use probability distributions to describe the properties of the effective scatterers, and then fundamental laws of wave propagation are applied to model the propagation process. The 3rd Generation Partnership Project (3GPP) has introduced the Spatial Channel Model (SCM) as a geometry-based model \cite{3gpp}. 
Later it was extended to the spatial channel model extended (SCME) \cite{baum2005interim}, where SCME supports wider bandwidth. In the Wireless World Initiative New Radio (WINNER) projects, the model was adapted to 15 different scenarios. Also, 3GPP has produced a 3-dimensional channel model 3D SCM for long term evolution advanced (LTE-advanced) \cite{thomas20133d}.

NGSM approaches focus on the paths between the transmitter and the receiver, and scatterers are not modeled explicitly. Tapped Delay Line (TDL) based models are the most known models of this type, where the propagation channel is considered as a number of delay taps. Each tap has a few characteristics, such as the average power, excess delay, and amplitude distribution function.

All of the aforementioned traditional models have limitations. As mentioned before, high complexity and site-specificity are the limitations of deterministic approaches. Furthermore, stochastic approaches are parametric and based on mathematical expressions that impose some assumptions on the model, which are not necessarily correct. In addition, some complicated mediums such as underwater acoustic channels or in-body channels may introduce some complex distortion effects, which are impossible to be expressed analytically. So, the need for a framework with the capability of learning complex models without any presumptions arises.

Recently, deep learning approaches have been applied in many areas, including communications. Examples in communications are CSI (Channel State Information) feedback compression and reconstruction \cite{wen2018deep}, joint source-channel coding \cite{farsad2018deep}, channel estimation and signal detection \cite{ye2018power} and many other applications. 


In this paper, in the context of deep learning approaches for communication applications, a novel channel modeling technique is introduced. This model overcomes some of the limitations of traditional methods. The idea is based on using generative models as intelligent frameworks for modeling the statistical distribution of channel measurements. Generative models are able to learn the distribution of the training data and generate new samples having the same statistical properties as the training data. Generative Adversarial Networks (GANs), as one of the most important types of generative models, have shown very good performances in modeling the distribution of the images. If somehow, the propagation channel can be considered as an image, GANs can be applied to model the distribution of channels as well. For this purpose, in this work, the 2D time-frequency responses of the channel are considered as images, which are called channel images. The main idea of this paper is to model the distribution of these channel images using GANs.
Furthermore, it is assumed that the user carrying the receiver is moving at various speeds.
\color{black} The effect of user speeds on channels is considered. We aim to propose a model that considers the user speed as an auxiliary parameter to generate channels matching the properties of a particular environment. 

A two-phase procedure is followed for the modeling approach. First, The distribution of channel images all having a reference user speed is modeled using Deep Convolutional GANs. In the second phase, we take advantage of StarGAN as an image-to-image translation technique to change the generated channel images with respect to the desired user speed. The model is applied to three different simulated channel types and also real measurement data. We first evaluate the performance of the model using common metrics such as Level Crossing Rate (LCR) and Average Fade Duration (AFD). These metrics have some shortcomings in evaluating the dissimilarity between different channel types, i.e., with these metrics, some channels of different types might show high similarity. To obtain a more reliable criterion, a new metric based on the Cepstral Distance Measure (CDM) of the mean of the autocorrelation of the data is introduced. This metric captures the specific frequency patterns of the time-frequency response for each channel type. It is used to evaluate the statistical similarity between measurement and generated data of each type.

The main contributions of this paper can be summarized as follows:
\begin{enumerate}
	\item Proposing a channel modeling technique which can generate channels with the desired user speed, having a similar distribution with the measurement data.
	\item Introducing a metric for statistical evaluation of the proposed model (checking the statistical analogy between the generated channels and the measurement data.)
\end{enumerate} 

This paper is organized as follows. Section \Romannum{2} provides background on the propagation channel, GANs, and image-to-image translation techniques. In Section \Romannum{3} problem definition is provided, in Section \Romannum{4} the channel modeling procedure is described, in Section \Romannum{5} experimental results are provided, and finally Section \Romannum{6} concludes the paper.

\section{Background}
\subsection{Propagation Channel}
Propagation channel is the medium between the transmitter and the receiver which affects the amplitude and the phase of the transmitted signal. The complex channel gain of multi-path fading channels can be modeled as a stochastic process as: 

\begin{equation}\mu (t)=\sum_{n=1}^{N} c_{n}e^{j(2\pi f_{n} t +\theta _{n})} ,\label{eq1}\end{equation}
where N is the number of paths, $c_{n}$, $f_{n}$ and $\theta_{n}$ are the gain, Doppler frequency and phase of the $n$th path, respectively.
The Short-Time Fourier Transform (STFT) of \eqref{eq1}, is used for a time-frequency representation of the channel response.

\subsection{Generative Adversarial Networks}
\label{ssec:GAN}
Machine learning models can be divided into two general categories: discriminative models and generative models. Discriminative models try to find a relation between inputs and outputs and usually are used in regression, prediction, and classification applications. Generative models, on the other hand, have the ability to learn the distribution of the data and can generate samples having a similar distribution with the training data. One of the most popular generative models is GAN \cite{goodfellow2014generative}.

The idea behind GAN is to train two networks simultaneously. A generator network has to learn the distribution of training data and generate samples having the same distribution with the training data from input noise. On the other hand, a discriminator network has to find the probability that a sample comes from the training data. At the first steps of the training, generated samples are almost noise, so it is easy for the discriminator to discern real data from generated data. As the training process goes on, the generator and the discriminator evolve in an adversarial process until the samples generated by the generator are so similar to the real samples that the discriminator can not distinguish generated data from the real ones. At this point, the generator has learned the distribution of the training data and can be used to generate new samples.

GANs have been used widely in various applications and are extended for example to Deep Convolutional GANs (DCGANs) \cite{radford2015unsupervised}, Conditional GANs \cite{gauthier2014conditional}, Boundary Equilibrium GANs (BEGANs) \cite{berthelot2017began}, and Wasserstein GANs (WGANs) \cite{gulrajani2017improved}.
\subsection{Image-to-Image Translation}
Image-to-image translation is mapping an input image in one domain to a corresponding image in another domain and is classified into two main categories: supervised and unsupervised. In supervised methods, corresponding pairs of images in two domains are available. In \cite{isola2017image} the authors use conditional adversarial networks for this purpose. In the unsupervised image-to-image translation, there are two sets of images, one consisting the input images and the other consisting the corresponding images. No paired images are available to know how one image is translated to its corresponding output such as in \cite{zhu2017unpaired,tang2019dual}.

All of the aforementioned image-to-image translation methods are efficient for translating images in two different domains. For more than two domains, they have to train different networks for each of the two domains. Authors in \cite{choi2018stargan} presented the first framework for multi-domain image-to-image translation, which uses only one generator for translating images in all different domains. They name their network StarGAN.   
\begin{figure*}[t]
	\centering
	\centerline{\includegraphics[width=1.2\columnwidth]{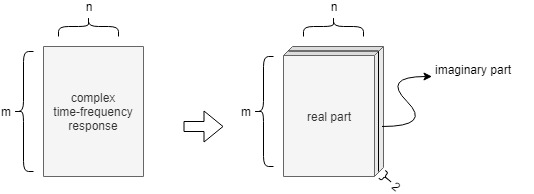}}
	\caption{Converting the channel time-frequency response into a channel image.}
	\label{channelAsImage}
\end{figure*}

\section{Problem Definition}

\subsection{Channel As an Image}
\label{ssec:chimag}
In Multiple-Input Multiple-Output (MIMO) systems, the transmitter/receiver antenna space is considered as the channel matrix($H$). Each element $H_{ij}$ in the channel matrix represents the channel between the $i^{th}$ transmitter antenna and the $j^{th}$ receiver antenna. In the MIMO settings, $H_{ij}$ usually is considered as a single value, and so $H$ will be a $r\times t$ matrix (number of receive antennas$\times$number of transmit antennas), and in some previous studies such as \cite{he2018deep}, $H_{ij}$ is treated as an image.

For a more detailed look at the channel, the single value of $H_{ij}$ can be substituted by a matrix that has $m\times n$ complex values, each value representing the effect of channel on a particular frequency and particular time ($m$ represents the number of frequency sub-carriers, and $n$ is the number of time slots).

In this work, we focus on a single-input single-output (SISO) transmission. 
The grid of the channel time-frequency response (between a pair of transmitter and receiver antennas) is translated into a $m\times n\times 2$ image by placing the real values in the first channel of image and the imaginary values in the second channel of it. Figure {\ref{channelAsImage}} shows how this translation is done for a grid of size $m\times n$. As mentioned in Section \ref{sec:introduction}, we denote these images by channel image. 
\begin{figure*}[ht]
	\centerline{\includegraphics[width=1.5\columnwidth]{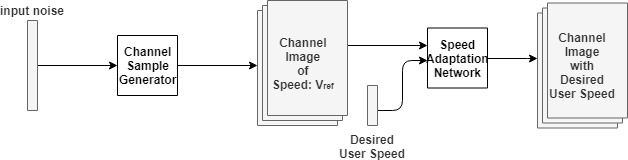}}
	\caption{The generation procedure of channel images with desired user speed.}
	\label{generateSpeed}
\end{figure*}

\begin{figure*}[ht]
	\centerline{\includegraphics[width=1.5\columnwidth]{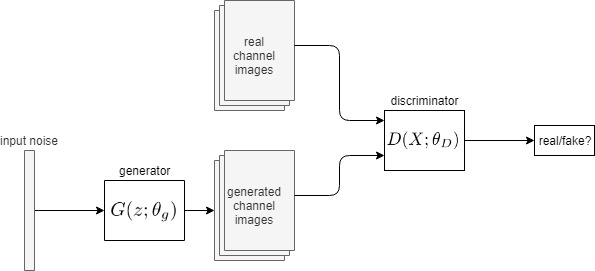}}
	\caption{Training procedure of DCGAN for learning the distribution of the channel images.}
	\label{DCGANTrain}
\end{figure*}

\subsection{Channel Modeling}
\label{ssec:ChannelModeling}
In data-driven channel modeling, which is followed in this paper, one has a set of collected channel measurements and is looking for a model that can find out the main statistics or features of the measurement data. With this model, new samples can be generated with the same statistics of the channel. Furthermore, there are a few parameters such as the user speed which in some-form affect the statistics of the channel. We intend to build a model that can incorporate such parameter (here we focus on user speed) when it generates a sample for that environment.

As for a formal problem definition, assume that many measurements of one particular channel type, i.e., a specific environment (like an urban area) are given. Each channel measurement is performed at a particular user speed. So we have a set of channel measurements $X_{Urban}=\left\{X_{1},X_{2},...,X_{n}\right\}$ containing $n$ samples of the channel measurements and a set of possible user speeds $V=\left\{v_{1},v_{2},...,v_{m}\right\}$ where $m$ is the number of distinct user speeds. Each $X_{i}$ is collected at a certain speed of the user. \textit{Note that} when the environment changes (channel type changes), another set of measurements like $X_{Rural}$ is obtained, where its elements are drawn from different statistics, and thus one needs to build another model for that environment.

The aim is to propose a model which can learn the statistical distribution of the measurement data of different channel types, and generate new samples for that environment. The set of generated samples ($\hat{X}$), should have the same distribution as the measurement data ($X$). The model should also capture the effect of different user speeds on channel measurements and generate channels with respect to a specific desired user speed. 

For evaluating the performance of such model, a metric is needed to evaluate how the samples in $\hat{X}$ emulate the statistics of the samples in $X$. Since we are talking about statistics of two sets, measures like Euclidean distance of samples do not work, instead, one should employ metrics that: \textit{1)} can demonstrate that the samples of the model trained using dataset $X$ are statistically similar to the samples in $X$ (the actual measured data),\textit{2)} can evaluate \textit{the dissimilarity} between the statistical distribution of the samples of the model trained using dataset $X$ and the samples of another set like $Y$,  measured for another channel type/user-speed.

For this purpose, there are a few of such metrics in literature, but 
as will be discussed later, they fail to simultaneously possess the two properties, especially, since we are dealing with two-dimensional samples (like channel images). In this study, thus, we have to introduce a proper comparison metric as well. 

\section{Data Driven Channel Modeling Procedure}

As discussed in Section \ref{ssec:chimag}, we can consider the channel measurements as channel images. With this idea, the problem of generating samples with the same distribution of channel measured samples can be posed as a problem in which, one is trying to generate images that have a similar shape to the images that are already in the training dataset.
This view of the problem inspired us to try to use generative models (briefly discussed in Section \ref{ssec:GAN}) as a tool for propagation channel modeling.

Generative models have many different variants. Among them, as we have different channel types and user speeds, the natural choice probably would be Conditional GAN (CGAN) as a type of GAN that is used for the labeled categorical dataset. CGAN considers the labels as the condition parameter and generates samples based on this parameter. We tried to train a CGAN to generate samples like the measurement data while considering the user speed as the CGAN conditions. Unfortunately, probably due to the high complexity of the samples, the designed CGAN did not converge properly, and the generated samples did not have similar desired statistics.

To manage the problem, we have proposed a two-phase solution. Especially, instead of trying to train a model that can \textit{directly} give us a channel sample \textit{for a particular channel type} and \textit{desired user-speed}, we generate the desired sample using two networks. This procedure is shown in Figure {\ref{generateSpeed}} pictorially, where:
\begin{itemize}
	\item \textbf{Channel sample generator:} In this network, the generator does not need to deal with different user speeds. The goal of the first network is to generate samples that have the same distribution as the measured data, assuming that all of them are collected at a nominal speed, $v_{ref}$. 
	\item \textbf{Speed adaptation network:} In the second step, the network modifies the generated sample such that its statistics match the statistics of the channel at the desired user speed, $v$.  
\end{itemize}

In the following two sections, the structures of the \textit{Channel sample generator} and \textit{Speed adaptation network} are described. Then, in Section \ref{ssec:eval_metric}, the metrics employed to evaluate the similarity between the generated and actual channel samples are discussed.

\subsection{Channel Sample Generator}

As for the first step, we need to train a network that learns the distribution of channel images (collected at $v_{ref}$) and generates similar samples. In this study, a DCGAN (briefly reviewed in Section \ref{ssec:GAN}) operates as the channel sample generator.

The training procedure is shown in Figure {\ref{DCGANTrain}}. For each environment, the set of collected channel measurements (e.g., $X_{Rural}$ or $X_{Urban}$) is considered as the training data. The generator tries to generate channel images having the same distribution with the training data i.e., it tries to find a mapping $G(z;\theta_{g})$ from the noise space to the data space, where $\theta_{g}$ is the set of parameters for $G$. The input noise is sampled from a prior distribution $P_{z}(z)$ that is usually considered a uniform distribution on [-1,1]. 
The discriminator network, denoted by $D(X;\theta_{D})$, tries to distinguish real channel images from the ones generated by the generator network. 

Both the generator and the discriminator are deep convolutional networks. In the training of the network, labels $0$ (representing fake images) are allocated to the generated channel images and labels $1$ (representing real images) are allocated to the real channel images. The discriminator performs a classification between real and generated channel images. The loss function is:
\begin{multline}
min_{G} max_{D} V(G,D) = E_{x\sim p_{data}(x)} [log D(x)]+ \\E_{z\sim p_{z}(z)} [log[1-D(G(z))]].\label{eq2}
\end{multline}
where the discriminator's output $D(x)$ is a single scalar representing the probability of $x$ coming from the measured dataset. During the training, the discriminator tries to maximize the loss function and classify its input images as real or fake (generated by the generator). On the other hand, The generator aims to fool the discriminator by minimizing $V(G,D)$. 

Obviously, in the first iterations of training, the generated images are almost noise, and it is easy for the discriminator to distinguish them from real channel images. As the training goes on, the generator improves in competition with the discriminator until the channel images generated by the generator are so similar to the training data that the discriminator cannot distinguish them. At this point, the generator learns the distribution of channel images and the network converges.

\subsection{Speed Adaptation Network}

\begin{figure}[h]
	\centering
	\begin{adjustbox}{minipage=\linewidth,scale=1}
		\begin{subfigure}{.49\textwidth}
			\centering
			\includegraphics[width=\linewidth]{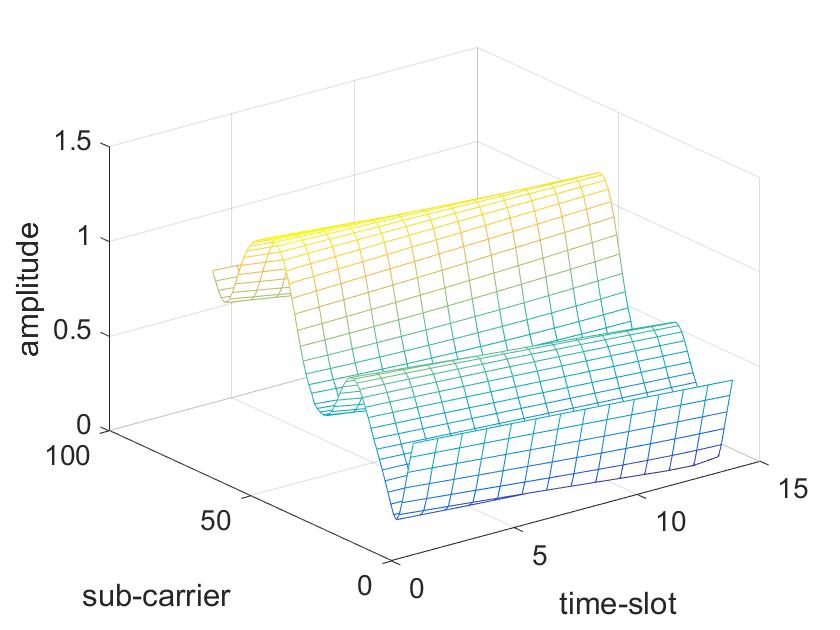}  
			\caption{User Speed = 25 km/h}
			\label{ETU_speed:sub-first}
		\end{subfigure}
		\begin{subfigure}{.49\textwidth}
			\centering
			\includegraphics[width=\linewidth]{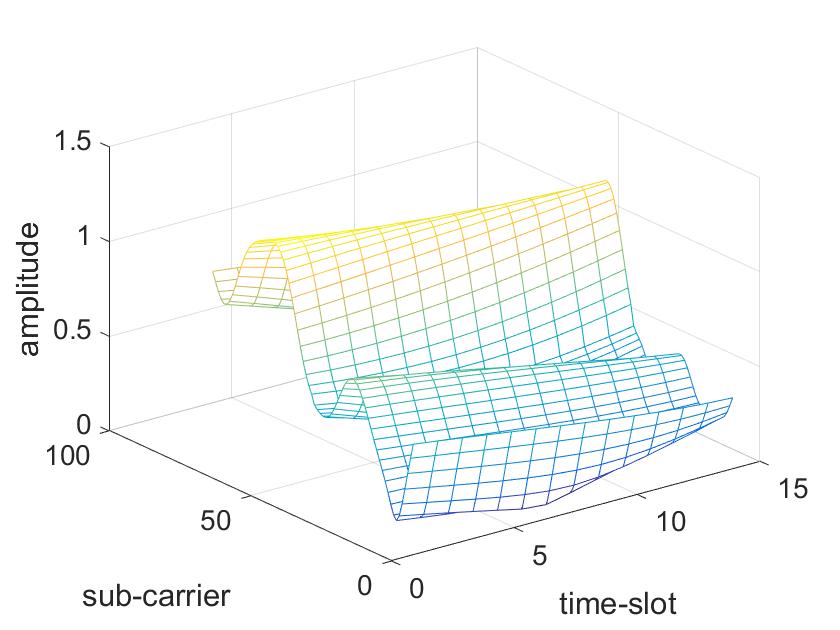}  
			\caption{User Speed = 50 km/h}
			\label{ETU_speed:sub-second}
		\end{subfigure}
		\begin{subfigure}{.49\textwidth}
			\centering
			\includegraphics[width=\linewidth]{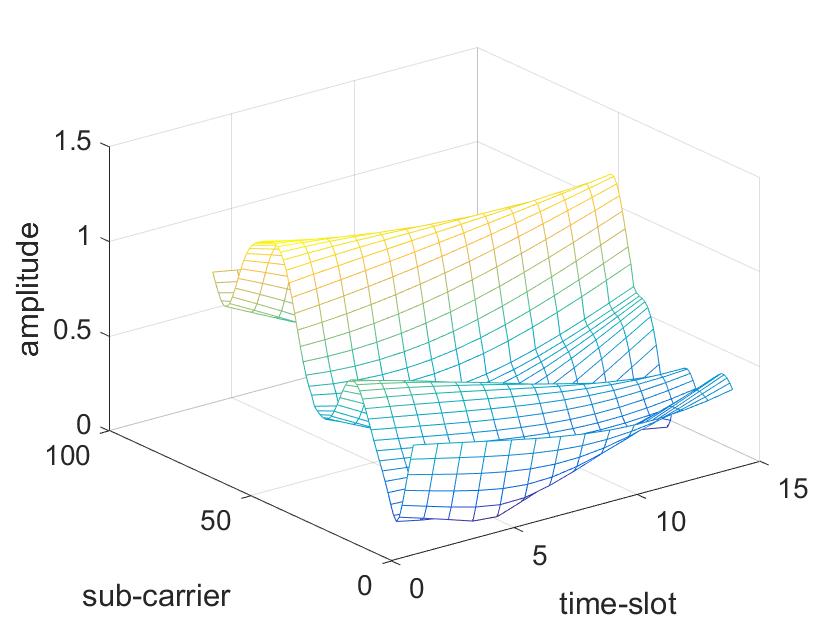}  
			\caption{User Speed = 75 km/h}
			\label{ETU_speed:sub-third}
		\end{subfigure}
		\begin{subfigure}{.49\textwidth}
			\centering
			\includegraphics[width=\linewidth]{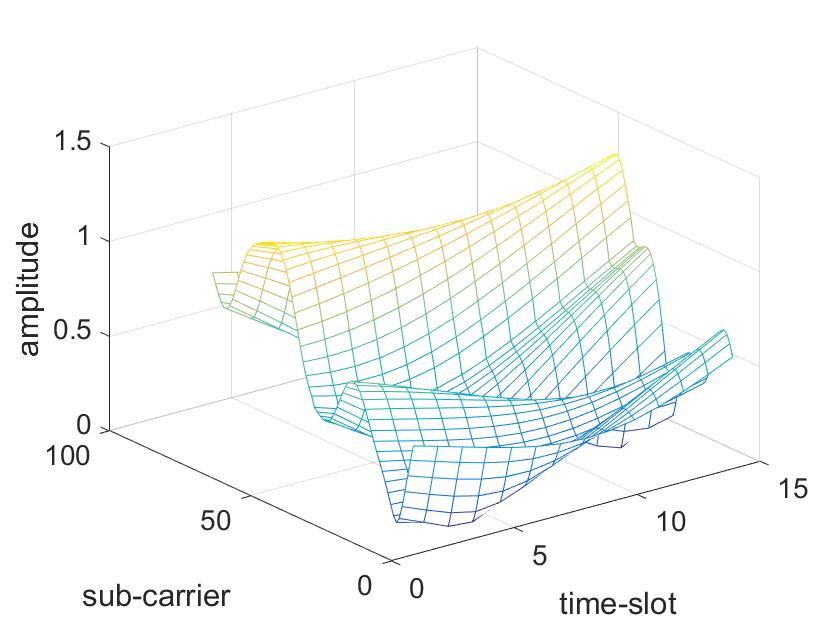}  
			\caption{User Speed = 100 km/h}
			\label{ETU_speed:sub-fourth}
		\end{subfigure}
		\caption{A sample of ETU channel with four different User Speeds: (\subref{ETU_speed:sub-first}) 25 km/h, (\subref{ETU_speed:sub-second}) 50 km/h, (\subref{ETU_speed:sub-third}) 75 km/h, (\subref{ETU_speed:sub-fourth}) 100 km/h.}
		\label{ETU_speed} 
	\end{adjustbox}
\end{figure}

User speed mainly affects the Doppler frequency, and its impact is seen in the time domain of the time-frequency response of the channel. Figure {\ref{ETU_speed}} shows a sample ETU channel with four different user speeds: $25, 50, 75 ,$ and $100$ km/h. As can be seen, when the user speed increases, the variations in time axis is increased. 

In the second step of the network, we aim to adjust the channel images such that the network learns how different user speeds affect the channel image and incorporates the effect of the user speed on the generated sample images.

One immediate idea to build such model is to train a deep net that gets channel images of $v_{ref}$ as input, and \textit{the corresponding channel image of \textbf{the same} sample when the user speed is, for example, $v_1$} as output. Training the network with many of these samples, one can hope that it will do the correct modification on the generated sample to convert them from $v_{ref}$ to the desired speed.

\textbf{The main challenge for this method} is that it is not possible to gather such dataset. We can definitely measure channel when the user is moving with speed $v_{ref}$ and $v_{1}$ but it is very hard to keep all the elements of environment the same between the two experiments. In other words, it is possible to have two sets of training data, one collected at $v_{ref}$ and the other collected at $v_{1}$, but it is not possible to determine which element of the first set is associated with which element of the second set.

\begin{figure*}[ht]
	\centerline{\includegraphics[width=1.2\columnwidth]{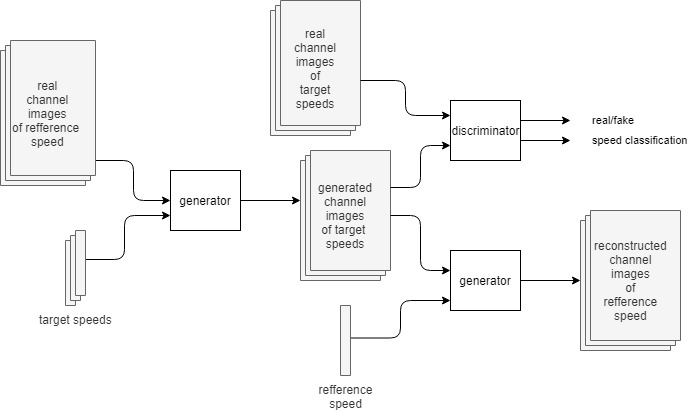}}
	\caption{Training procedure of StarGAN for converting channel images with user speed $v_{ref}$ to channel images with user speeds $v_{1}, v_{2},..., v_{n}$.}
	\label{starGANTrain}
\end{figure*}

In essence, we are dealing with two datasets, and we intend to learn how to translate from the first domain to the second domain (without knowing the exact match between the dataset members).  \textit{StarGAN} is a structure that has been suggested for such settings to translate between multiple domains. StarGAN accepts as inputs a reference channel image and a label (of the desired target domain) and generates an image associated with the reference domain but in the target domain. 
The training procedure, which is basically similar to the adversarial training of GANs, is shown in Figure {\ref{starGANTrain}}. 

For our goal, starGAN is trained by feeding the generator with  
reference channel images with user speed $v_{ref}$, and labels of the target user speeds $v_{1}, v_{2},..., v_{m}$. The generator tries to generate channel images of target user speeds. The generated channel images, along with real target channel images, are fed into the discriminator. The discriminator has two tasks: first, to distinguish fake channel images from the real ones, second, to classify the channel images based on their labels (user speeds). To make sure the procedure preserves the correspondence between the reference and the target images, StarGAN should also train the generator to reconstruct the reference channel image. This time, the generated channel image (at the target speed) and the reference speed are given to the generator as inputs, and it has to generate an image that should be as close as possible to the original reference channel images (at $v_{ref}$).

When training is complete, we use the generator of the StarGAN; the reference channel image with user speed $v_{ref}$ and the desired user speed are given to the generator. It, then, generates a channel with adjusted user speed based on the desired speed. 

\subsection{Evaluation metric}
\label{ssec:eval_metric}
As we mentioned in Section \ref{ssec:ChannelModeling}, commonly used metrics for evaluation of the performance of the channel models are looking at the 1D channel response (either in frequency or in time).

\begin{figure}[h]
	\centerline{\includegraphics[width=0.5\columnwidth]{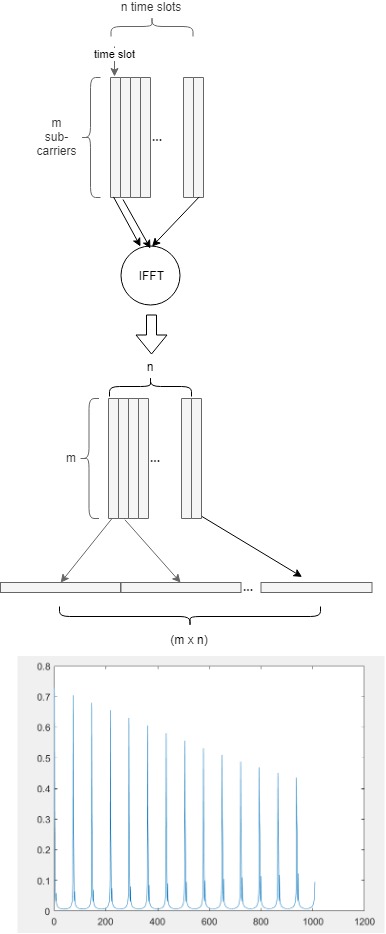}}
	
	\caption{The process of converting 2D time-frequency responses into a 1D sequence.}
	\label{1Dseq}
\end{figure}

In this work, we have also utilized these measures. Since the emphasis in this work is on generating 2-dimensional time-frequency responses of the channel, we have additionally presented a new metric to evaluate the pattern of the generated 2D images. 
The 2-dimensional ($m\times n$) time-frequency response provides frequency responses in consecutive time-slots. By putting along all the $n$ frequency responses horizontally (similar to the last step in Figure {\ref{1Dseq}}), we actually obtain a 1-dimensional quasi-periodic signal. Its first $m$ elements are corresponding to the first time-slot, its second $m$ elements are corresponding to the second time slot, and so on. This operation is repeated for all of the time-frequency responses. The resulting 1D sequences are denoted as $x_i[n]$ $(i = 1,2, ..., N)$, where $N$ is the number of samples (time-frequency responses). We take the autocorrelation of this quasi-periodic signal($x_i[n]$) as:

\color{black}

\begin{equation}R_{xx} (m)=E(x_i(n)x_i(n+m)) .\label{eq3}\end{equation}

To evaluate the similarity of the generated samples to the real measurements, one can consider the mean of the resulting autocorrelation functions over the samples. In Section \ref{ssec:Exp_pA}, few of these mean autocorrelation functions are depicted, where we can verify a good match between the generated and measured samples. 

Next we would like to obtain a quantitative comparison of the autocorrelation function results. To achieve this goal, we employ the real Cepstrom ($c[n]$) of the mean autocorrelation functions as:

\begin{equation}c[n]=\mbox{real}\left[\mbox{IFFT}\left[\log\left[\mbox{abs}\left[\mbox{FFT}\left[\mbox{mean}(R_{xx}[m])\right]\right]\right]\right]\right],\label{eq4}\end{equation} 

The reason we choose Cepstrom is that we are looking for repeated features. Moreover, Cepstrum does two operations on $c[n]$: 
1) \textbf{compression}, which makes the comparison tractable, and
2) \textbf{spectral smoothing}. Autocorrelation holds spectral information with respect to the Wiener-Kinchin theorem \cite{yates2014probability} that provides the relation between
the power spectrum and the autocorrelation function. By taking Cepstrom of the autocorrelation, we smooth the spectrum, skip the redundant information, and keep the envelope which plays the main role for our comparison.

The similarity measure now can be defined as the Mean Squared Error (MSE) between the lower Cepstral coefficients for generated and measurement data.

\begin{figure*}[ht]
	\centerline{\includegraphics[width=1.5\columnwidth]{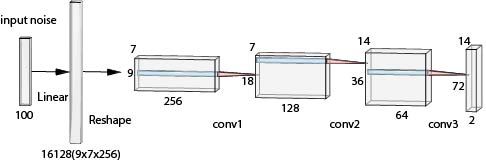}}
	\caption{Structure of DCGAN's generator network for channel modeling.}
	\label{genDCGAN}
\end{figure*}
\begin{figure*}[t]
	\centerline{\includegraphics[width=1.5\columnwidth]{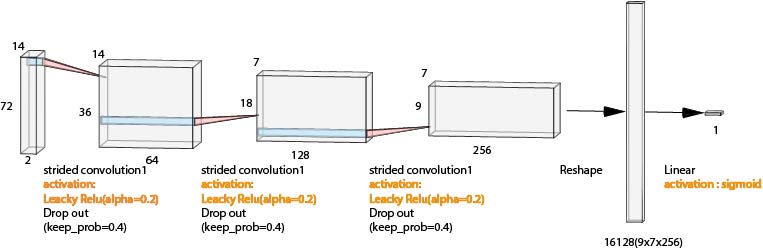}}
	\caption{Structure of DCGAN's discriminator network for channel modeling.}
	\label{discDCGAN}
\end{figure*}

\section{Experimental Results}

In this section, we describe the exact structure of the deep networks used for \textit{Channel Sample Generator} and \textit{Speed Adaptation Network}. We also discuss the training procedure and the results obtained from each network. The quality of the generated samples is then compared based on commonly used criteria as well as the introduced metrics based on the Cepstrum of the autocorrelations.

\subsection{Channel Sample Generator}
\label{ssec:Exp_pA}

\subsubsection{\textbf{Network Structure}}
DCGANs of the Channel Sample Generator uses deep convolutional networks for both the discriminator and the generator of DCGAN. The structure of the generator and the discriminator networks are depicted in Figure {\ref{genDCGAN}} and Figure {\ref{discDCGAN}}, respectively. 

The network is trained for 10 epochs for the total of 40000 channel images with bach size of 64. The size of the noise vector is 100, the learning rate is $0.0002$, and Adam optimizer with $\beta_{1}=0.5$ is used for optimization.

\begin{figure}[ht]
	\centering
	\begin{adjustbox}{minipage=\linewidth,scale=1}
		\begin{subfigure}{.49\textwidth}
			\centering
			\includegraphics[width=\linewidth]{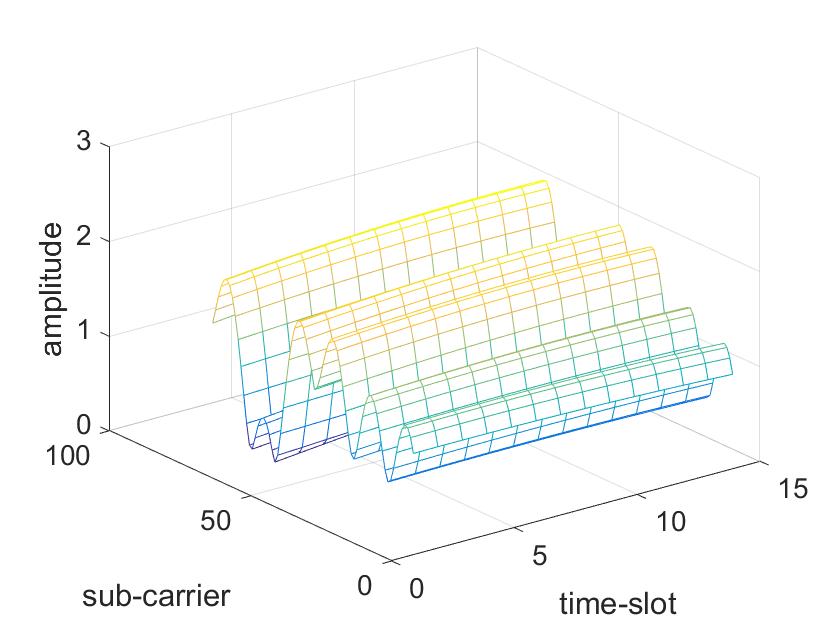}  
			\caption{ETU}
			\label{Training_Channels:sub-first}
		\end{subfigure}
		\begin{subfigure}{.49\textwidth}
			\centering
			\includegraphics[width=\linewidth]{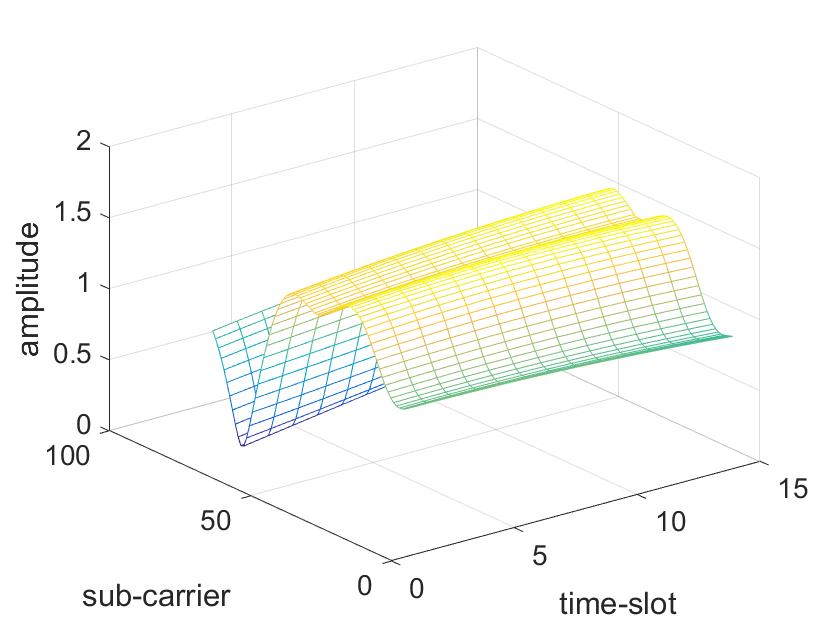}  
			\caption{EVA}
			\label{Training_Channels:sub-second}
		\end{subfigure}

		\begin{subfigure}{.49\textwidth}
			\centering
			\includegraphics[width=\linewidth]{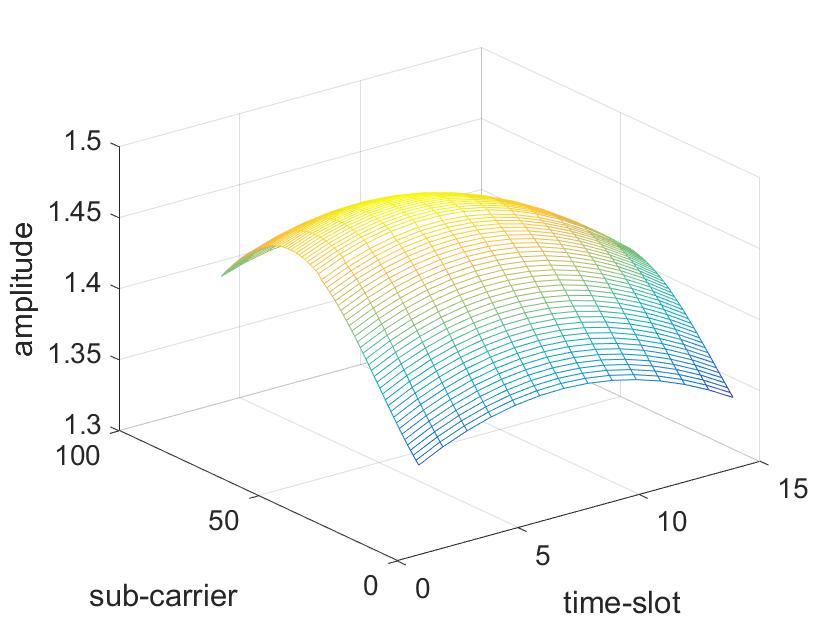}  
			\caption{PedA}
			\label{Training_Channels:sub-third}
		\end{subfigure}
		\begin{subfigure}{.49\textwidth}
			\centering
			\includegraphics[width=\linewidth]{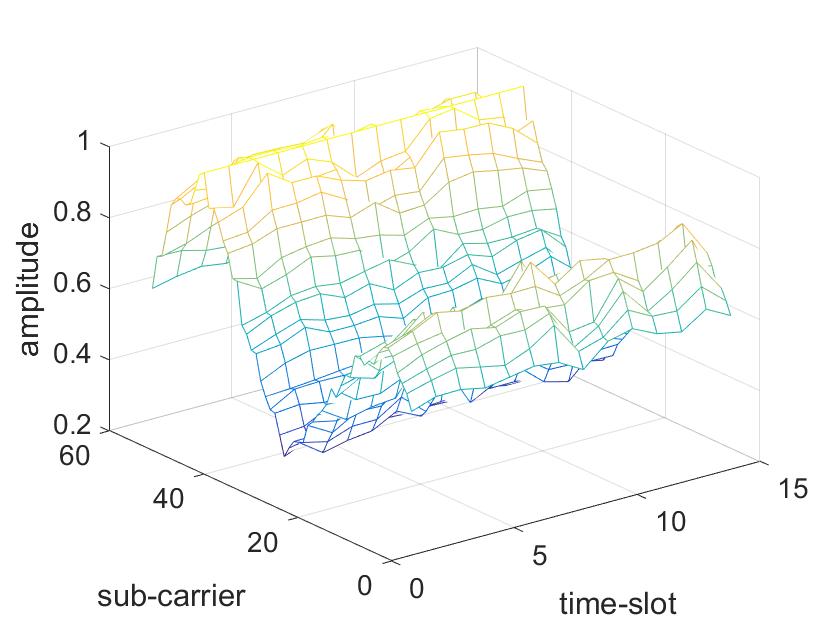}  
			\caption{Experimental Data}
			\label{Training_Channels:sub-fourth}
		\end{subfigure}
		\caption{A random sample of training data for (\subref{Training_Channels:sub-first}) ETU, (\subref{Training_Channels:sub-second}) EVA, (\subref{Training_Channels:sub-third}) PedA simulated channels, and (\subref{Training_Channels:sub-fourth}) experimental data.}
		\label{Training_Channels}
	\end{adjustbox}
\end{figure}

\begin{figure}[ht]
	\centering
	\begin{adjustbox}{minipage=\linewidth,scale=1}
		\begin{subfigure}{.49\textwidth}
			\centering
			\includegraphics[width=\linewidth]{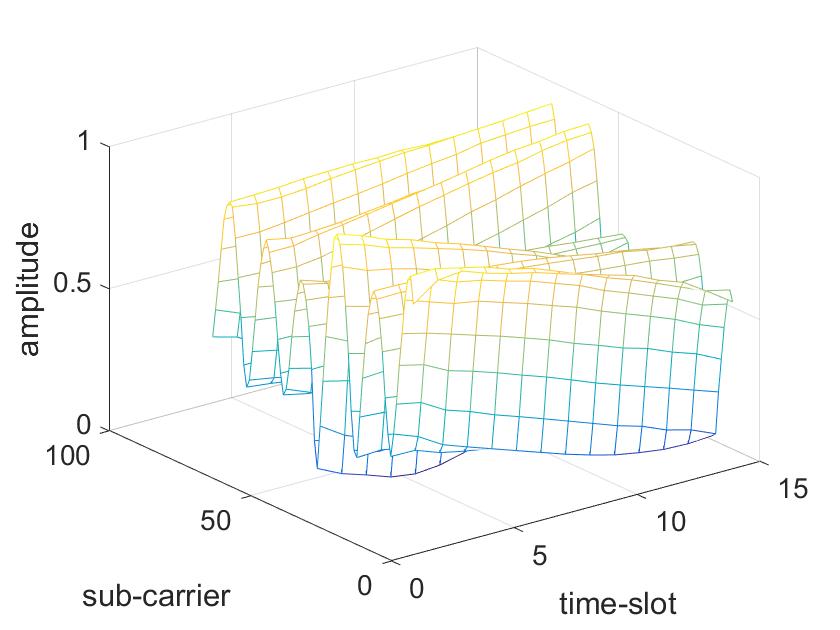}  
			\caption{ETU generated}
			\label{generatedChannelsDCGAN:sub-first}
		\end{subfigure}
		\begin{subfigure}{.49\textwidth}
			\centering
			\includegraphics[width=\linewidth]{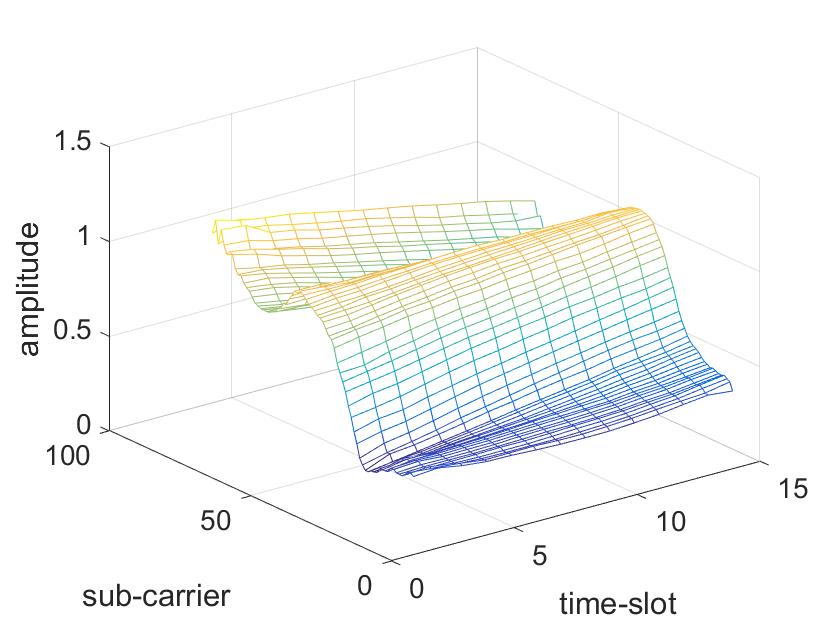}  
			\caption{EVA generated}
			\label{generatedChannelsDCGAN:sub-second}
		\end{subfigure}

		\begin{subfigure}{.49\textwidth}
			\centering
			\includegraphics[width=\linewidth]{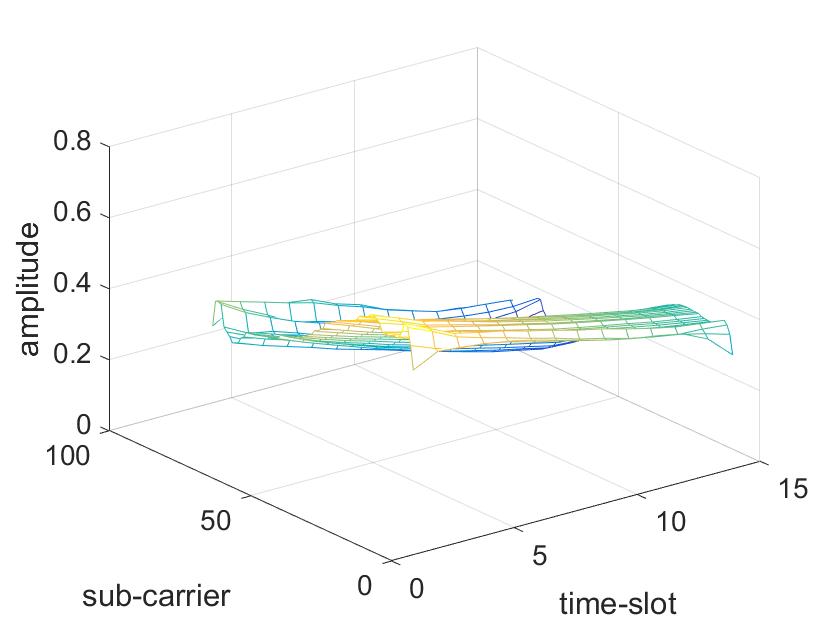}  
			\caption{PedA generated}
			\label{generatedChannelsDCGAN:sub-third}
		\end{subfigure}
		\begin{subfigure}{.49\textwidth}
			\centering
			\includegraphics[width=\linewidth]{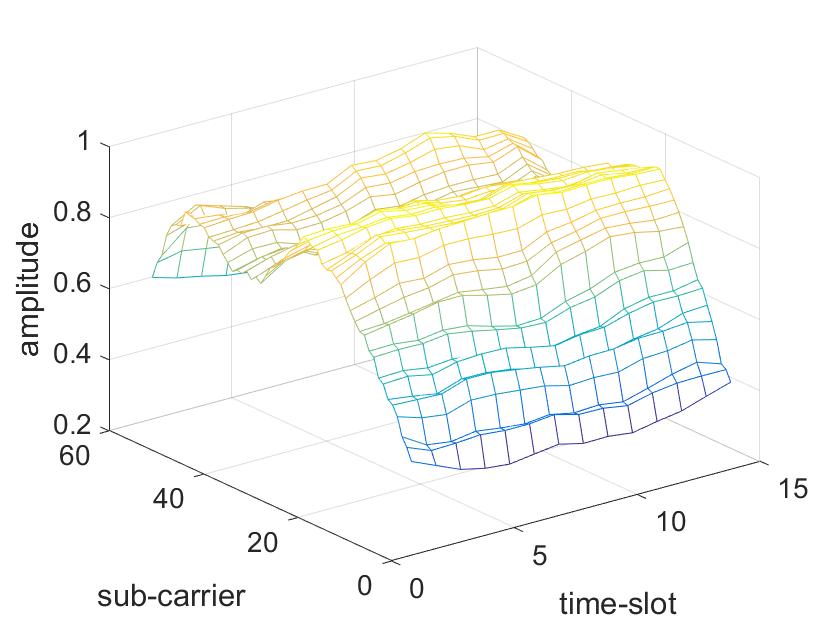}  
			\caption{Experimental Data generated}
			\label{generatedChannelsDCGAN:sub-fourth}
		\end{subfigure}
		\caption{A random sample of generated channels for (\subref{Training_Channels:sub-first}) ETU, (\subref{Training_Channels:sub-second}) EVA, (\subref{Training_Channels:sub-third}) PedA, and (\subref{Training_Channels:sub-fourth}) experimental data.}
		\label{generated_Channels}
		\label{generatedChannelsDCGAN}
	\end{adjustbox}
\end{figure}

\begin{figure*}[t]
	\centering
	\begin{adjustbox}{minipage=\linewidth,scale=0.8}
		\begin{subfigure}{.33\textwidth}
			\centering
			\includegraphics[width=\linewidth]{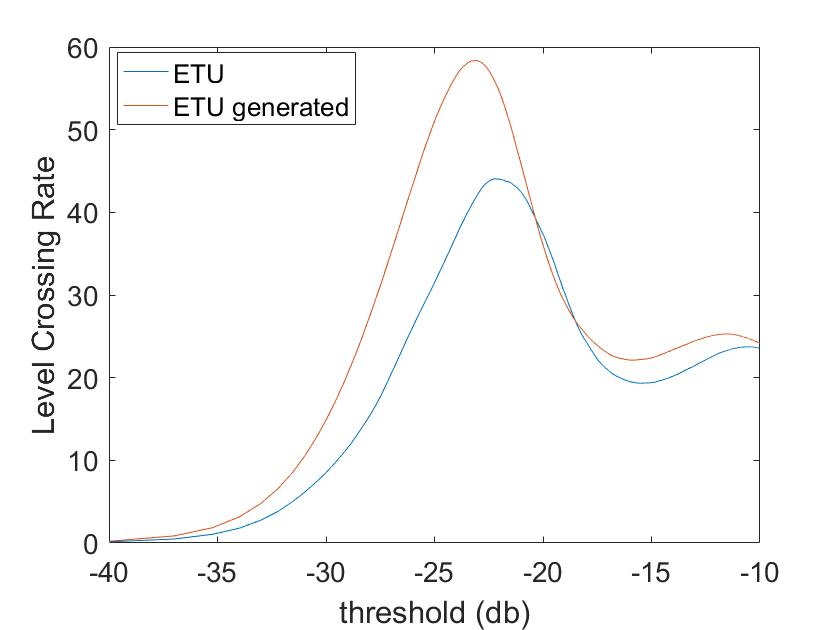}  
			\caption{}
			\label{lcr:sub-first}
		\end{subfigure}
		\begin{subfigure}{.33\textwidth}
			\centering
			\includegraphics[width=\linewidth]{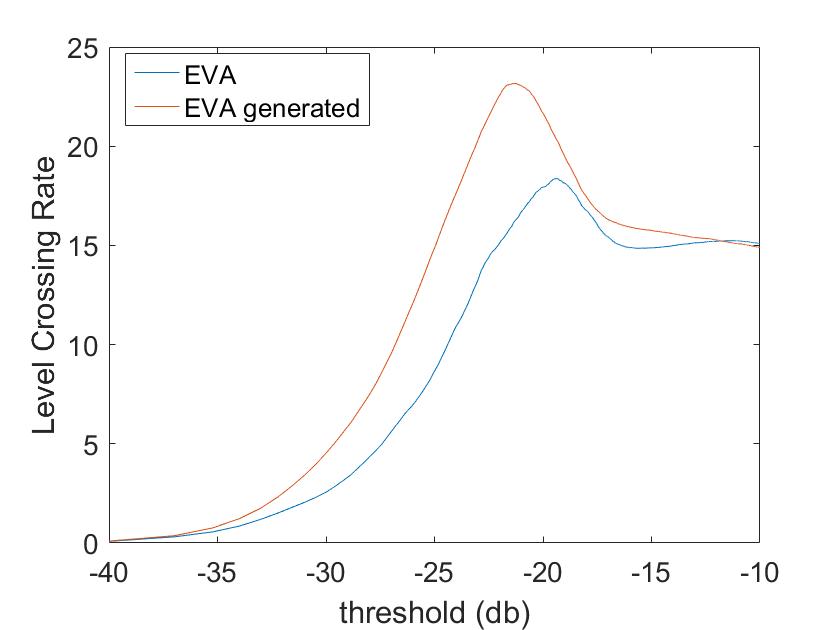}  
			\caption{}
			\label{lcr:sub-second}
		\end{subfigure}
		\begin{subfigure}{.33\textwidth}
			\centering
			\includegraphics[width=\linewidth]{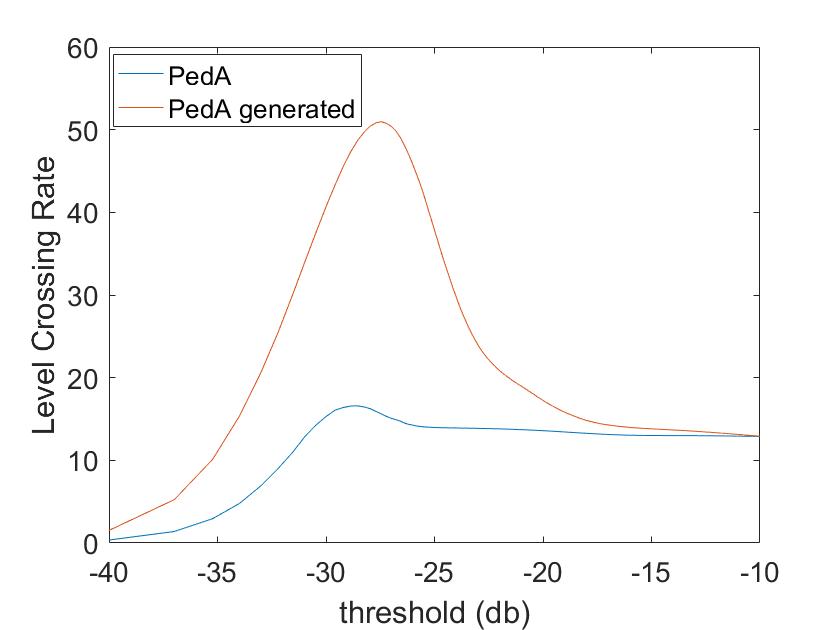}  
			\caption{}
			\label{lcr:sub-third}
		\end{subfigure}
		\caption{Level Crossing Rate (LCR) of measurement and generated channels for three channel types:(\subref{lcr:sub-first}) ETU, (\subref{lcr:sub-second}) EVA, and (\subref{lcr:sub-third}) PedA.}
		\label{lcr}
	\end{adjustbox}
\end{figure*}

\begin{figure*}[t]
	\centering
	\begin{adjustbox}{minipage=\linewidth,scale=0.8}
		\begin{subfigure}{.33\textwidth}
			\centering
			\includegraphics[width=\linewidth]{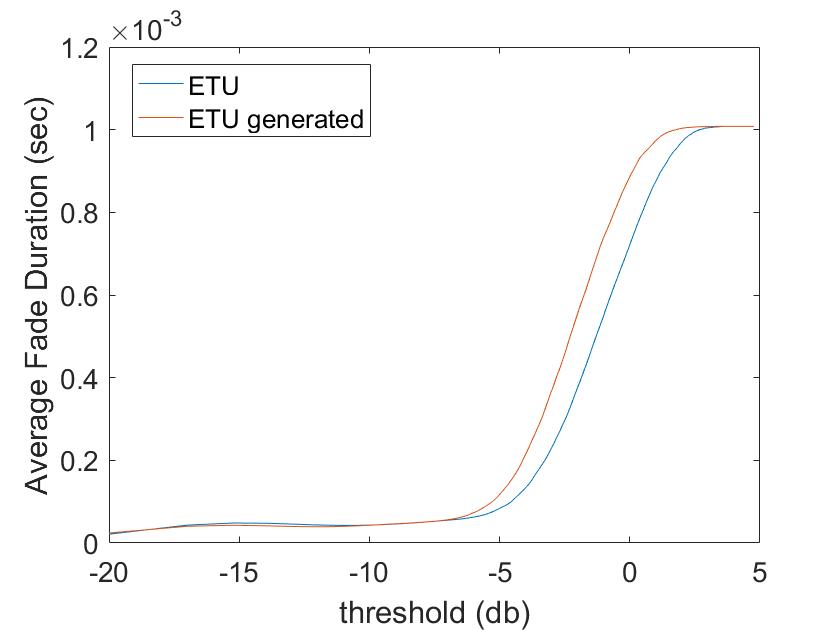}  
			\caption{}
			\label{afd:sub-first}
		\end{subfigure}
		\begin{subfigure}{.33\textwidth}
			\centering
			\includegraphics[width=\linewidth]{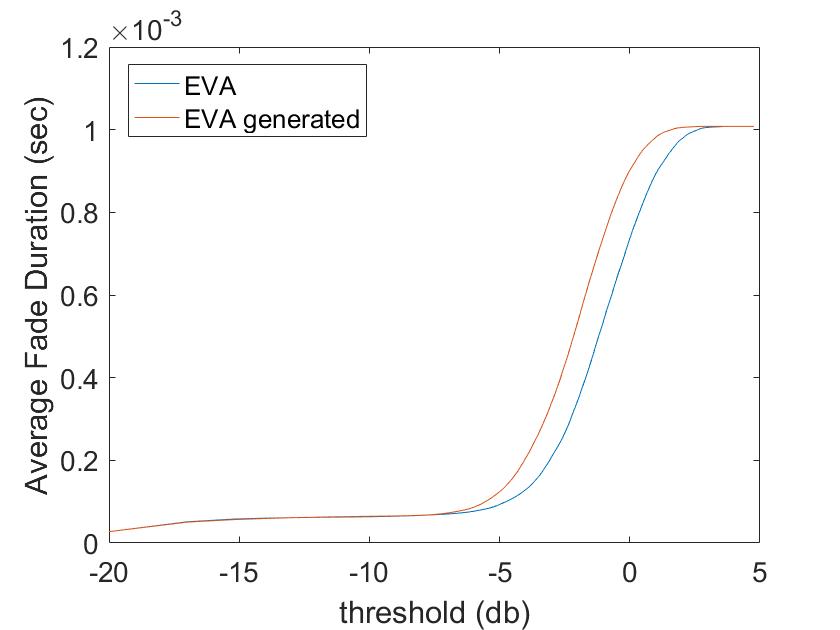}  
			\caption{}
			\label{afd:sub-second}
		\end{subfigure}
		\begin{subfigure}{.33\textwidth}
			\centering
			\includegraphics[width=\linewidth]{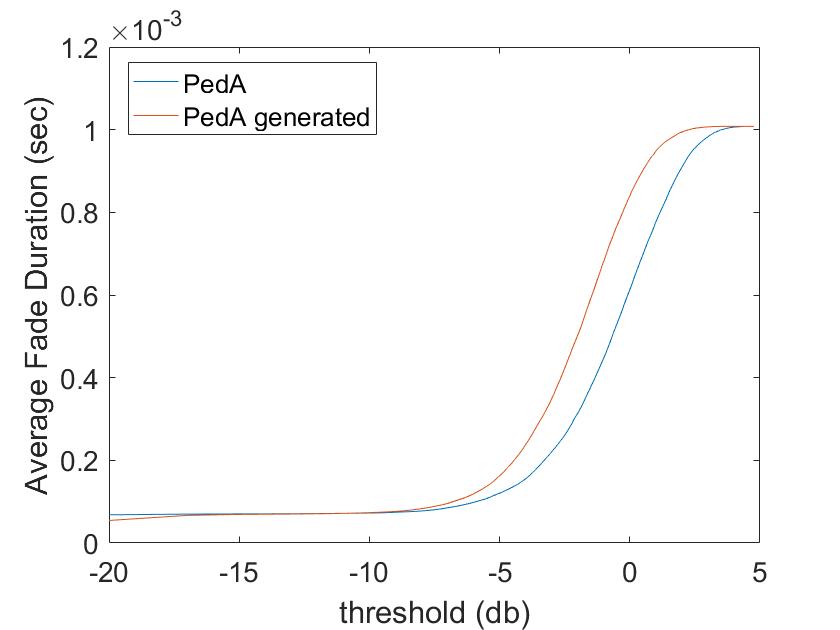}  
			\caption{}
			\label{afd:sub-third}
		\end{subfigure}
		\caption{Average Fade Duration (AFD) of measurement and generated channels for three channel types:(\subref{afd:sub-first}) ETU, (\subref{afd:sub-second}) EVA, and (\subref{afd:sub-third}) PedA.}
		\label{afd}
	\end{adjustbox}
\end{figure*}

\subsubsection{\textbf{Training Data}}
 
To train the model, we should collect some channel measurements from the environment. The model should provide us samples which have similar statistics. 

To evaluate the performance of our method, in this paper, we have collected the training data in two different experiments. 
\begin{itemize}
	\item In the first experiment (which is mainly for verification purposes), instead of the actual channel measurements, we employ a simulator to generate channel samples. The samples generated by this simulator are then considered as the actual channel measurements, and our model aims to generate samples similar to these (virtually) measured samples. 
	
	As we do not have the burden of actual channel measurements, with this method, we can easily simulate different environments and evaluate the performance of our model.
	
	\item In the second experiment, and for a realistic scenario, we collect some channel samples from the environment and then model our environment using the proposed scheme. 
	
\end{itemize}

For the first experiment, we use the Vienna link simulator \cite{rupp2016vienna} to generate our training data (time-frequency responses) for three types of channels: Extended Typical Urban model (ETU), Extended Vehicular A model (EVA) and Pedestrian A model (PedA). In terms of channel complexity, as shown in Figure {\ref{Training_Channels}}, ETU provides a more complex channel than EVA and EVA is more complicated than PedA, i.e., the variations in the frequency axis are more significant. 
We trained our channel sample generator to model each of these distributions separately.  

For each channel type, 40000 time-frequency responses of size $72\times 14$ are sampled as the training data. The simulation parameters are listed in Table\ref{simulationParameters}.
\begin{table}
	\centering
	\caption{Parameters for generating training data by Vienna LTE link simulator.}
	\label{simulationParameters}
	\begin{tabular}{|l|l|}
		\hline
		channel type & ETU/EVA/PedA  \\
		\hline
		Band width & 1.4 MHz \\
		\hline
		Simulation type & SUMIMO \\
		\hline
		Carrier frequency & 2.1 GHz \\
		\hline
		filtering & fast fading \\
		\hline
		user speed & ETU: 50,EVA: 80, PedA: 3 km/h \\
		\hline  
	\end{tabular}
\end{table}

For the second experiment, we use a commercially available Wi-Fi card equipped with Atheros AR9380 chipset \cite{Xie:2015:PPD:2789168.2790124} to collect data. Atheros AR9380 is able to report back the Channel State Information (CSI), and so it can be used as a cheap spectrum analyzer. 
For this experiment, the transmitter and the receiver are placed about 3m apart. We walked between them, moved the transmitter and the receiver to change the channel between them, and generate distinct measurements so that they can be used as training data. The channel bandwidth was 20 MHz. With this bandwidth, Atheros chipset reports the channel state for 56 sub-carriers. Thus, we will obtain 56 complex numbers in one CSI measurement for each transmission pair. We measured the CSI for a long time and selected 40000 time-frequency grids of size $56\times 14$ from it.

\begin{figure*}[t]
	\centering
	\begin{adjustbox}{minipage=\linewidth,scale=0.8}
		\begin{subfigure}{.33\textwidth}
			\centering
			\includegraphics[width=\linewidth]{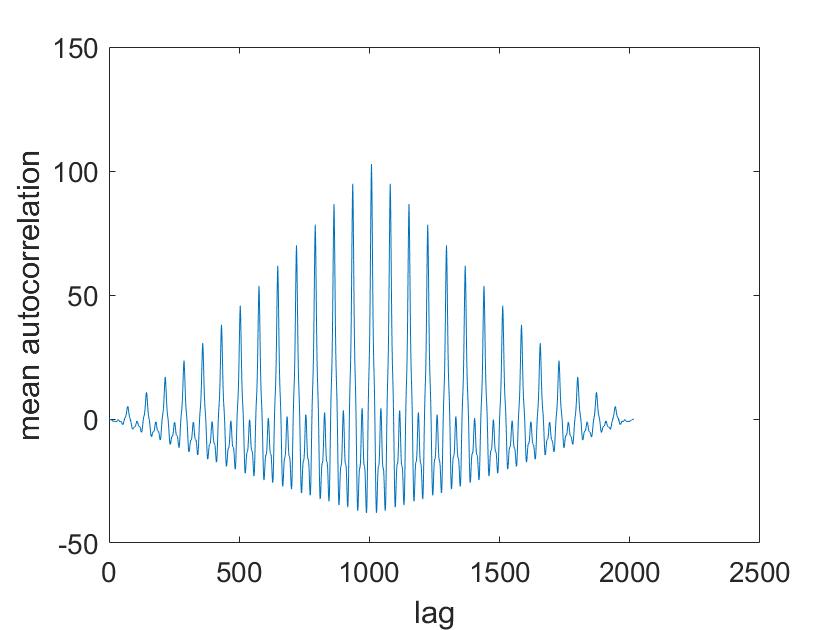}  
			\caption{}
			\label{xc_DCGAN:sub-first}
		\end{subfigure}
		\begin{subfigure}{.33\textwidth}
			\centering
			\includegraphics[width=\linewidth]{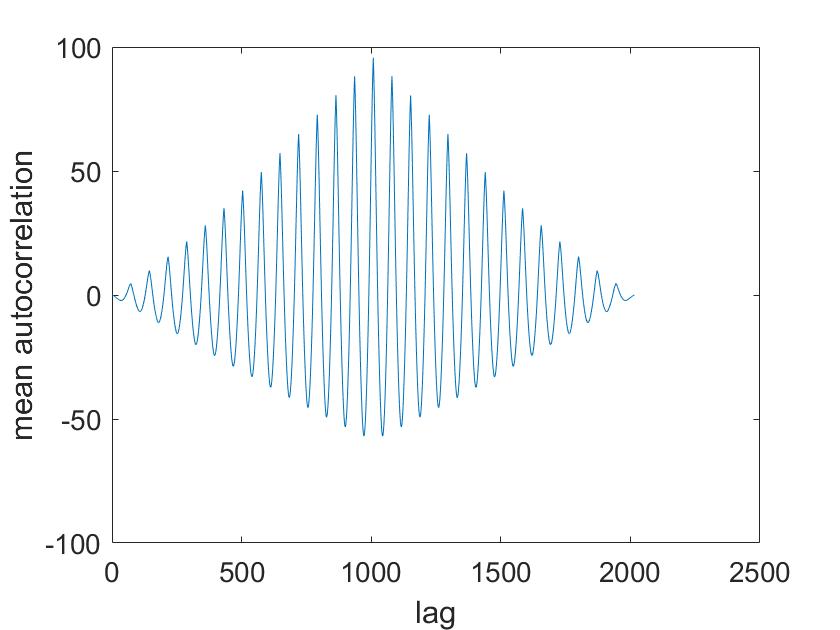}  
			\caption{}
			\label{xc_DCGAN:sub-second}
		\end{subfigure}
		\begin{subfigure}{.33\textwidth}
			\centering
			\includegraphics[width=\linewidth]{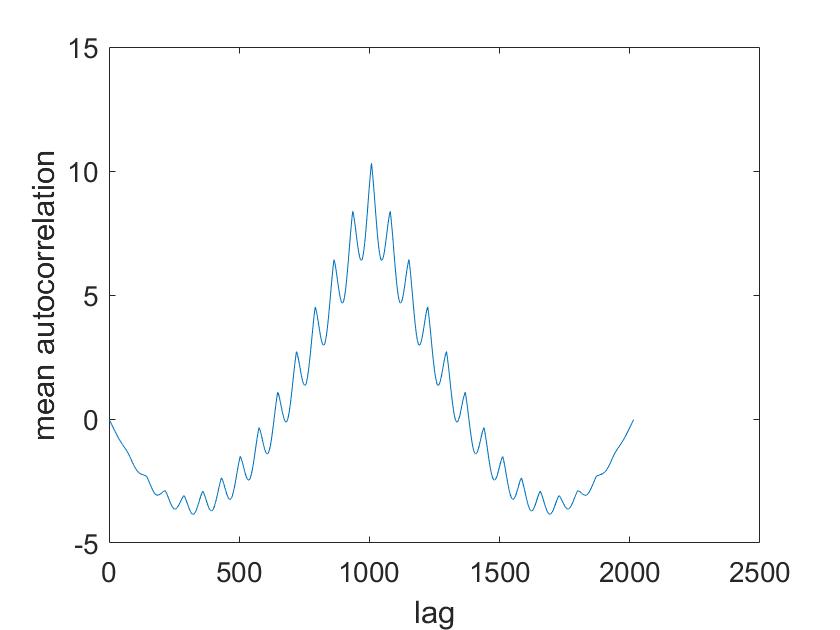}  
			\caption{}
			\label{xc_DCGAN:sub-third}
		\end{subfigure}
		\caption{The mean autocorrelation of the training data for three different channel types: (\subref{xc_DCGAN:sub-first}) ETU, (\subref{xc_DCGAN:sub-second}) EVA, and (\subref{xc_DCGAN:sub-third}) PedA.}
		\label{xc_DCGAN}
	\end{adjustbox}
\end{figure*}

\begin{figure*}[t]
	\centering
	\begin{adjustbox}{minipage=\linewidth,scale=0.8}
		\begin{subfigure}{.33\textwidth}
			\centering
			\includegraphics[width=\linewidth]{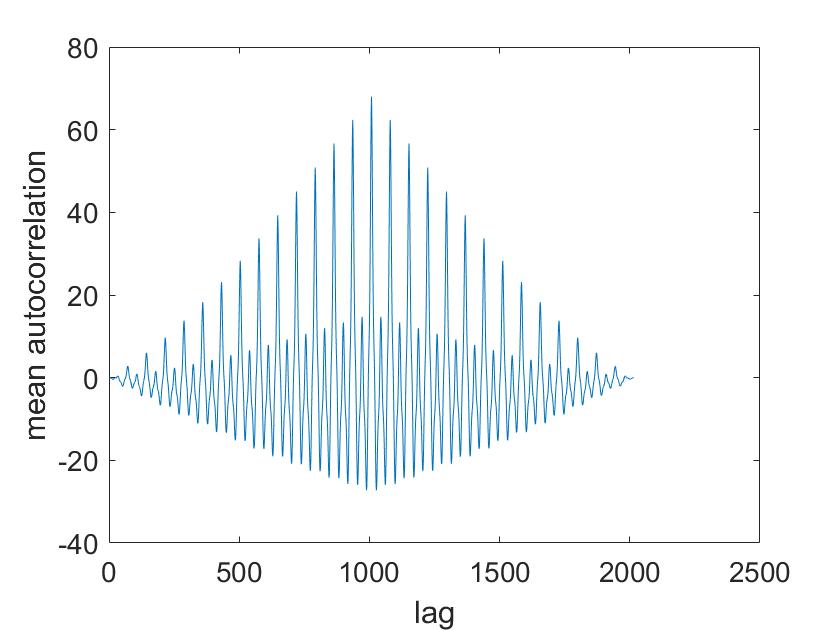}  
			\caption{}
			\label{xc_DCGAN_G:sub-first}
		\end{subfigure}
		\begin{subfigure}{.33\textwidth}
			\centering
			\includegraphics[width=\linewidth]{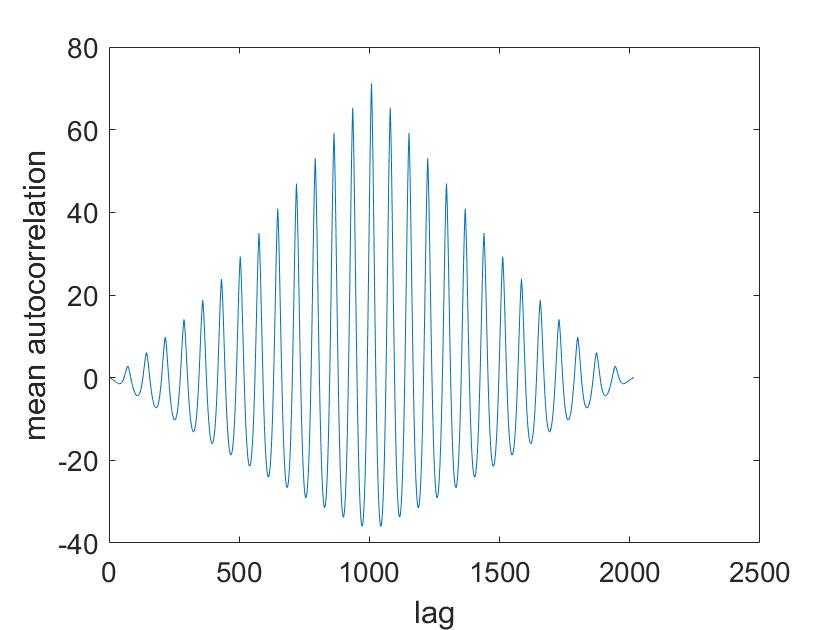}  
			\caption{}
			\label{xc_DCGAN_G:sub-second}
		\end{subfigure}
		\begin{subfigure}{.33\textwidth}
			\centering
			\includegraphics[width=\linewidth]{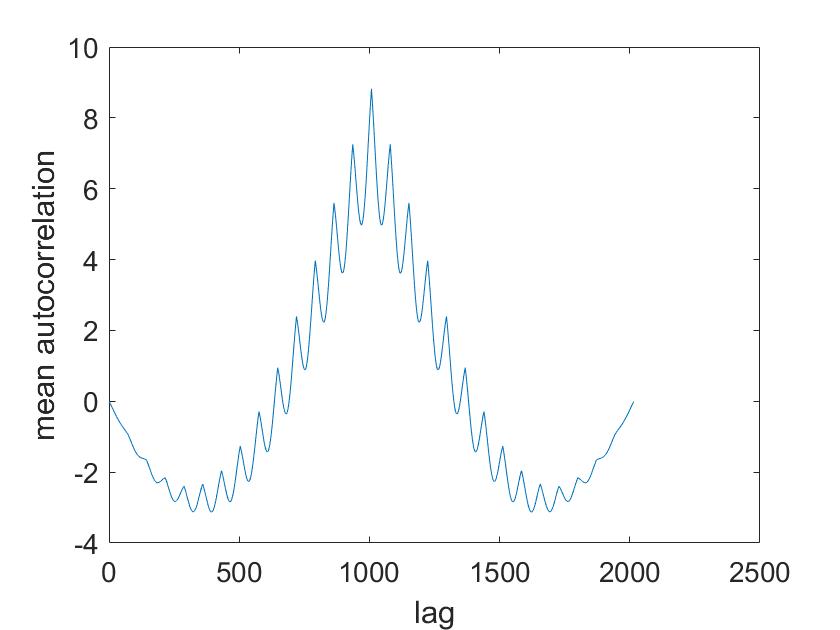}  
			\caption{}
			\label{xc_DCGAN_G:sub-third}
		\end{subfigure}
		\caption{The mean autocorrelation of the generated
			channels for three different channel types: (\subref{xc_DCGAN_G:sub-first}) ETU, (\subref{xc_DCGAN_G:sub-second}) EVA, and (\subref{xc_DCGAN_G:sub-third}) PedA.}
		\label{xc_DCGAN_G}
	\end{adjustbox}
\end{figure*}

\subsubsection{\textbf{Performance Analysis}}
In this section, we first present a few of samples of the generated time-frequency responses and then verify the statistical similarity of the generated time-frequency responses with the measurement data.

We train a model for each of the datasets, i.e., ETU, EVA, PedA, and real experimental data.  
Then we use the model to generator channel images from the learned distribution of the training data. The generated channel images are then used to obtain the complex time-frequency responses of the channel. 
The absolute value of samples of generated time-frequency responses for each case is shown in Figure {\ref{generatedChannelsDCGAN}}. By comparing Figure {\ref{generatedChannelsDCGAN}} with Figure {\ref{Training_Channels}}, visually, we can observe high analogy between the measured and generated data which verifies that they both come from a common distribution.

To show the good statistical closeness between the generated and actual samples, we use the commonly used metrics as well as the metric we introduced in Section \ref{ssec:eval_metric}.

First, we evaluate our model with the metrics of the Level Crossing Rate (LCR) and Average Fade Duration (AFD). LCR measures the number of times a signal exceeds a certain level, and it shows how fast the fading is for that channel. AFD measures the average time in which a signal is below a specific power. These metrics are used for 1-dimensional temporal data; however, the time-frequency responses are two-dimensional. So our two-dimensional time-frequency responses have to be converted into 1-dimensional temporal data. For this purpose, we take inverse fast Fourier transforms from each column of time slots and put along all the $n$ columns horizontally to have a 1D sequence. This process and a sample of these 1D sequences are illustrated in Figure {\ref{1Dseq}}. 

The charts of LCR and AFD for three different channel types are illustrated in Figure {\ref{lcr}} and Figure {\ref{afd}}, respectively. The LCR for  ETU and EVA channels demonstrate a good match in the location and amplitude of the pick. For PedA channel the location of picks is the same, but the amplitudes do not match. AFD for all channel types shows high similarity between measurement and generated channels.

Although the LCR and AFD metrics show a good match between the generated and the actual samples, they are not able to discriminate between different environments, even for the actual samples. For example, in Figure \ref{afd}, the AFD of the actual ETU and EVA samples are similar. 

The CDM in Section \ref{ssec:eval_metric}, though, is able to differentiate between different environments. As discussed before, for this metric, we first compute the mean autocorrelation function of the actual and generated channel samples. Figure {\ref{xc_DCGAN}} shows the results for ETU, EVehA, and PedA actual channel samples. In Figure {\ref{xc_DCGAN_G}} on the other hand, the mean autocorrelation of the generated channels is depicted. 
By comparing the two figures, it can be visually verified that the autocorrelation function of the generated channels of each type demonstrates a very high similarity to their corresponding autocorrelation function of the measured data. It can also be observed that different channel types resulted in different autocorrelation functions.

As mentioned in Section \ref{ssec:eval_metric}, to analyze beyond the visual comparison, we have used CDM for comparing the autocorrelation functions. 
The CDM for different channel types is listed in Table \ref{tableModel}. It is clear from the table that the distance between generated channels and their corresponding measurement is an order of magnitude less than the distance between generated channels of one type and measurements of different types.

\begin{table}
	\centering
	\caption{Cepstral Distance Measure for different channel types\\}
	\begin{tabular}{l|l|l|l}
		
		& \makecell{\\ ETU generated\\ \\} &  \makecell{\\ EVA generated\\ \\} &  \makecell{\\ PedA generated\\ \\}\\
		\hline
		\makecell{\\ ETU\\  \\} & $4.12\times 10^{-5}$ & $3.64\times 10^{-4}$ & $1.00\times 10^{-3}$ \\
		\hline
		\makecell{\\ EVA\\  \\} & $5.95\times 10^{-4}$ & $7.98\times 10^{-6}$ & $6.52\times 10^{-4}$ \\
		\hline
		\makecell{\\ PedA\\  \\} & $13.00\times 10^{-3}$ & $8.36\times 10^{-4}$ & $5.35\times 10^{-7}$
		
	\end{tabular}
	\label{tableModel}
\end{table}	

		\begin{figure*}[h]
	\centering
	\begin{adjustbox}{minipage=\linewidth,scale=0.8}
		\begin{subfigure}{.33\textwidth}
			\centering
			\includegraphics[width=\linewidth]{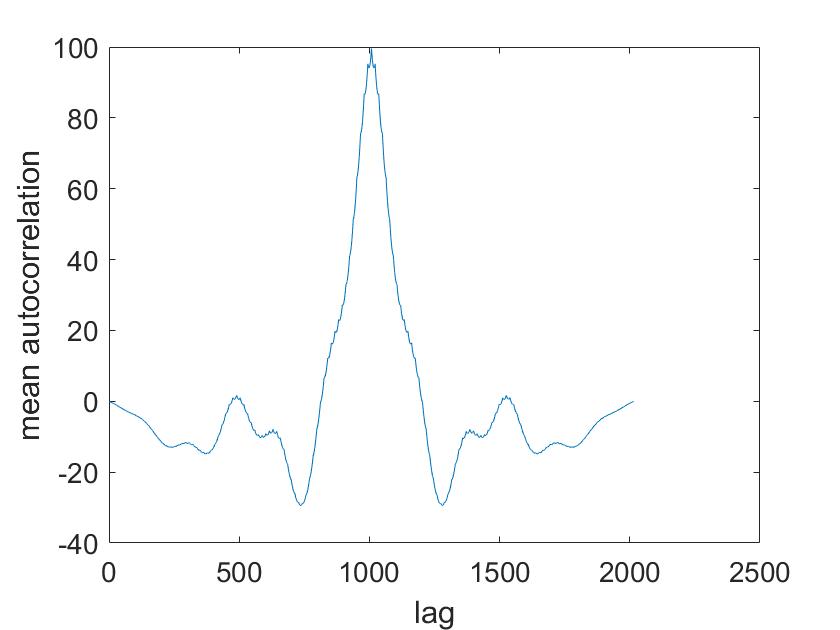}  
			\caption{}
			\label{xc_speed:sub-first}
		\end{subfigure}
		\begin{subfigure}{.33\textwidth}
			\centering
			\includegraphics[width=\linewidth]{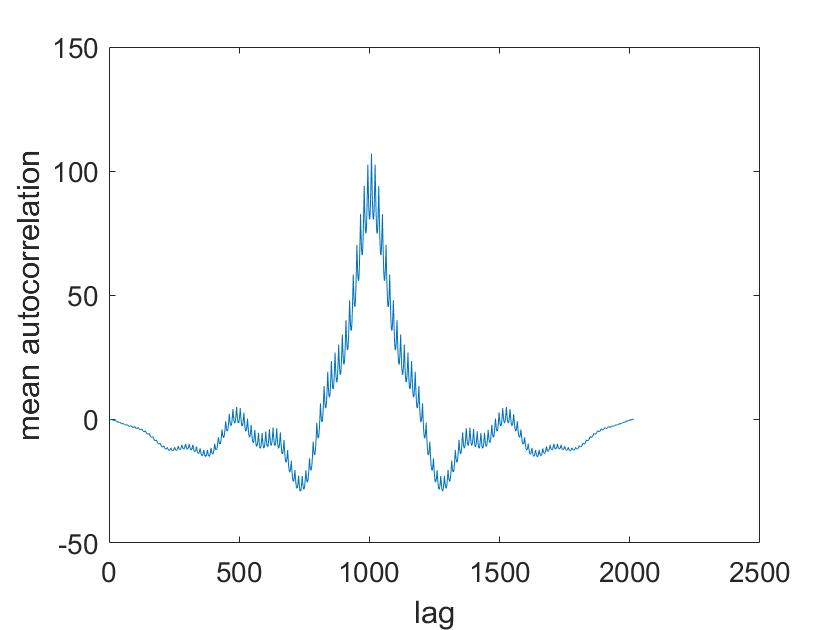}  
			\caption{}
			\label{xc_speed:sub-second}
		\end{subfigure}
		\begin{subfigure}{.33\textwidth}
			\centering
			\includegraphics[width=\linewidth]{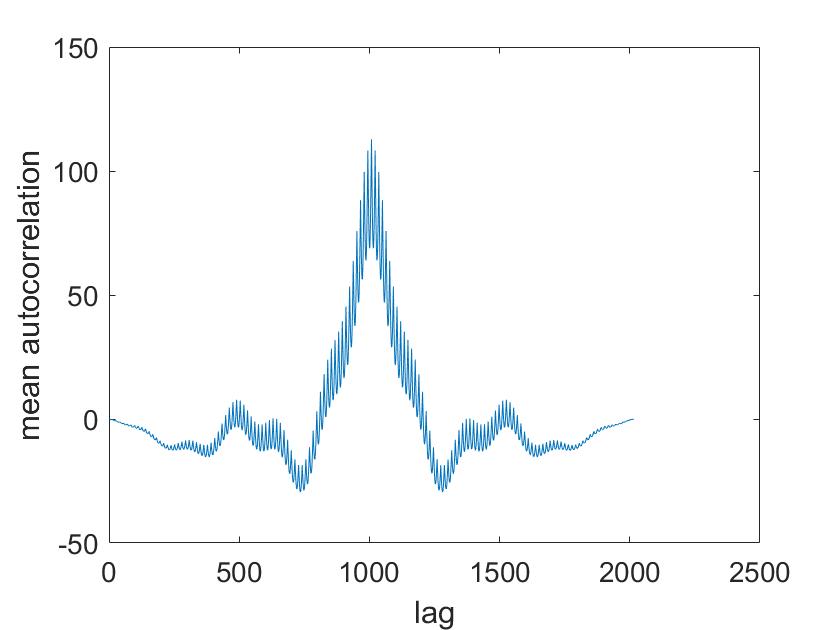}  
			\caption{}
			\label{xc_speed:sub-third}
		\end{subfigure}
		\caption{The mean autocorrelation of the actual ETU channels with three User Speeds: (\subref{xc_speed:sub-first}) 25 km/h, (\subref{xc_speed:sub-second}) 75 km/h, (\subref{xc_speed:sub-third}) 100 km/h.}
		\label{xc_speed}
	\end{adjustbox}
\end{figure*}

\begin{figure*}[h]
	\centering
	\begin{adjustbox}{minipage=\linewidth,scale=0.8}
		\begin{subfigure}{.33\textwidth}
			\centering
			\includegraphics[width=\linewidth]{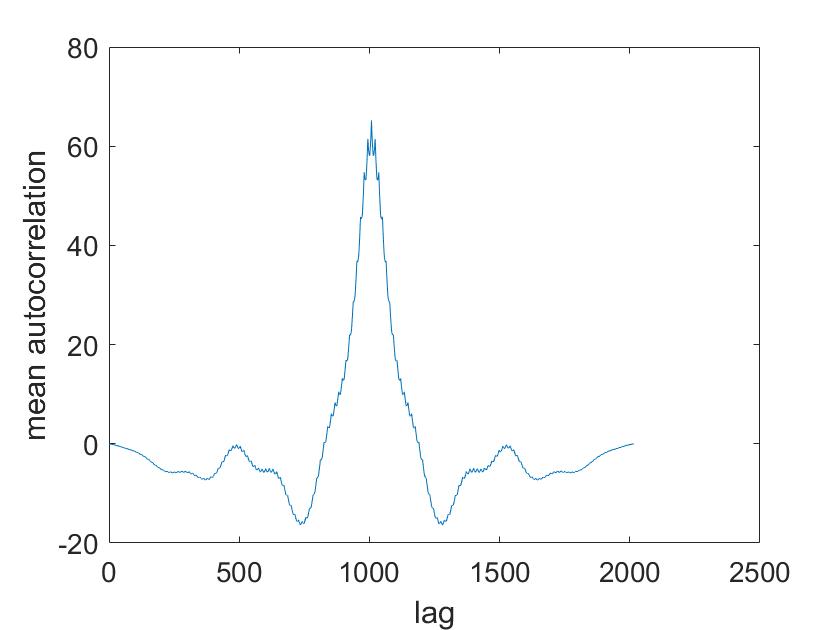}  
			\caption{}
			\label{xc_speed_G:sub-first}
		\end{subfigure}
		\begin{subfigure}{.33\textwidth}
			\centering
			\includegraphics[width=\linewidth]{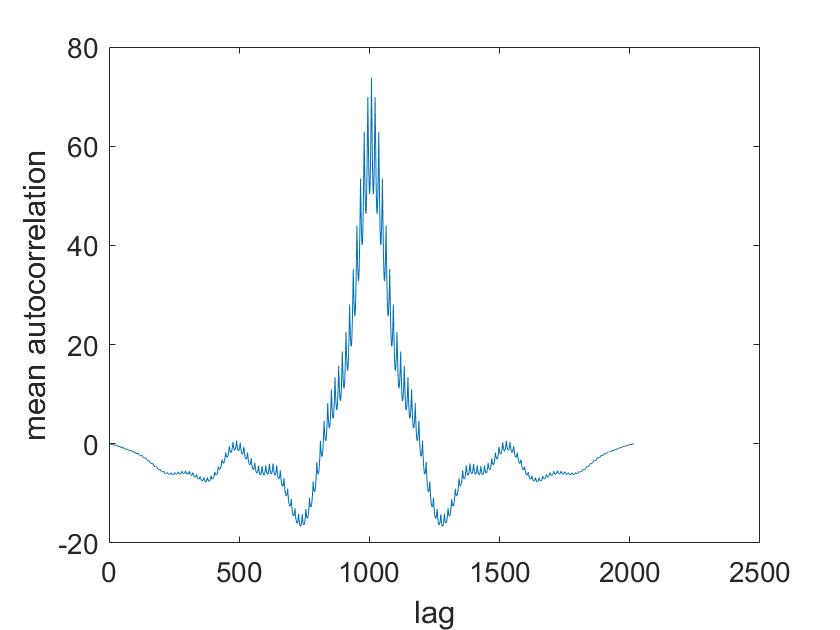}  
			\caption{}
			\label{xc_speed_G:sub-second}
		\end{subfigure}
		\begin{subfigure}{.33\textwidth}
			\centering
			\includegraphics[width=\linewidth]{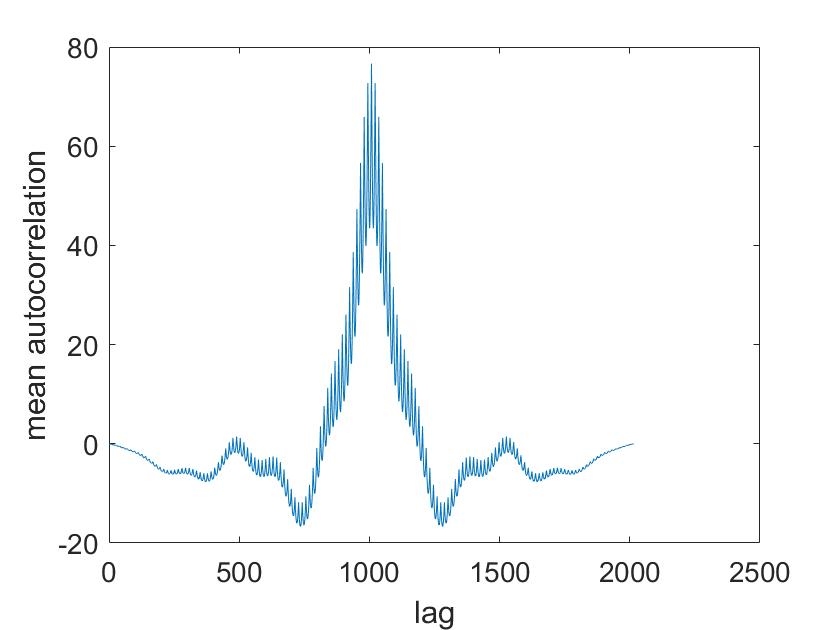}  
			\caption{}
			\label{xc_speed_G:sub-third}
		\end{subfigure}
		\caption{The mean autocorrelation of the generated ETU channels with three User Speeds: (\subref{xc_speed_G:sub-first}) 25 km/h, (\subref{xc_speed_G:sub-second}) 75 km/h, (\subref{xc_speed_G:sub-third}) 100 km/h.}
		\label{xc_speed_G}
	\end{adjustbox}
\end{figure*}

\begin{figure}[t]
	\centering
	\begin{adjustbox}{minipage=\linewidth,scale=1}
		\begin{subfigure}{.49\textwidth}
			\centering
			\includegraphics[width=\linewidth]{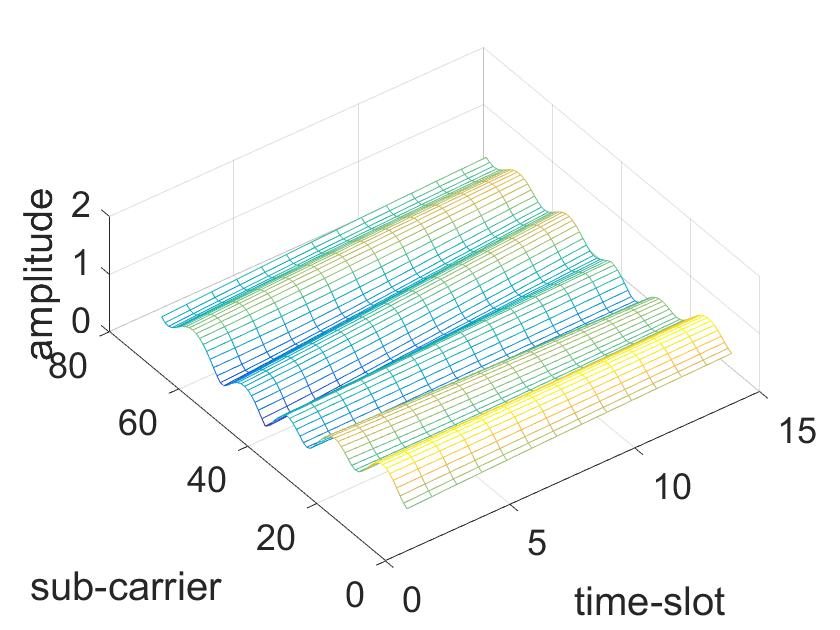}  
			\caption{}
			\label{Speed_G:sub-first}
		\end{subfigure}
		\begin{subfigure}{.49\textwidth}
			\centering
			\includegraphics[width=\linewidth]{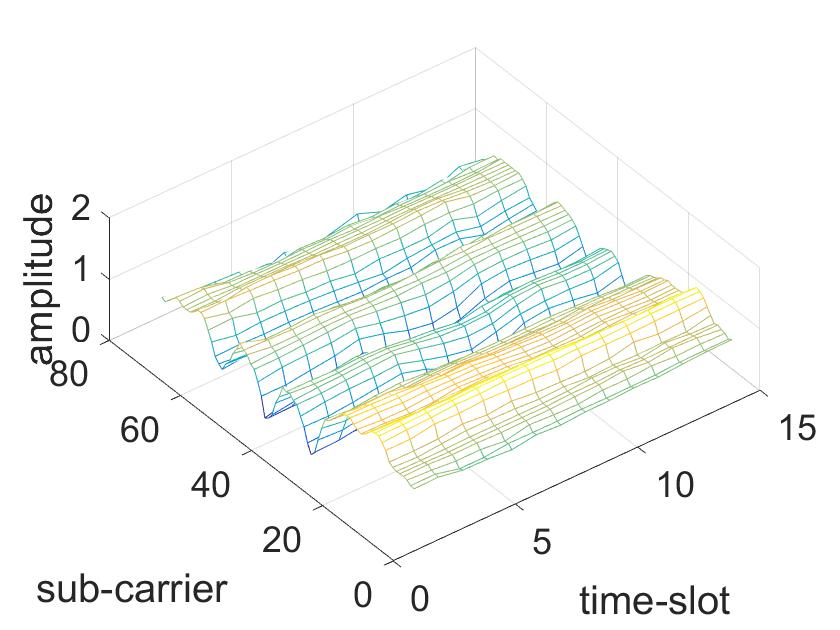}  
			\caption{}
			\label{Speed_G:sub-second}
		\end{subfigure}
		\begin{subfigure}{.49\textwidth}
			\centering
			\includegraphics[width=\linewidth]{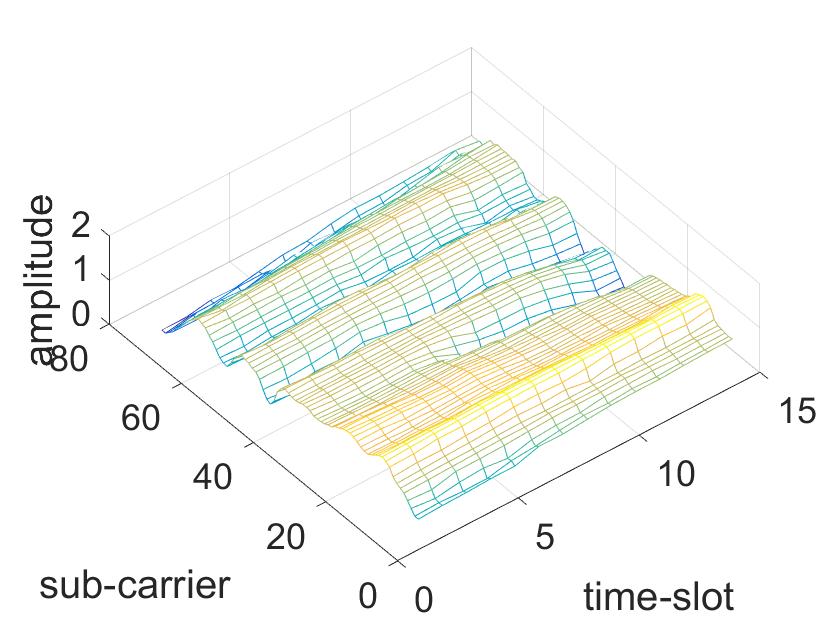}  
			\caption{}
			\label{Speed_G:sub-third}
		\end{subfigure}
		\begin{subfigure}{.49\textwidth}
			\centering
			\includegraphics[width=\linewidth]{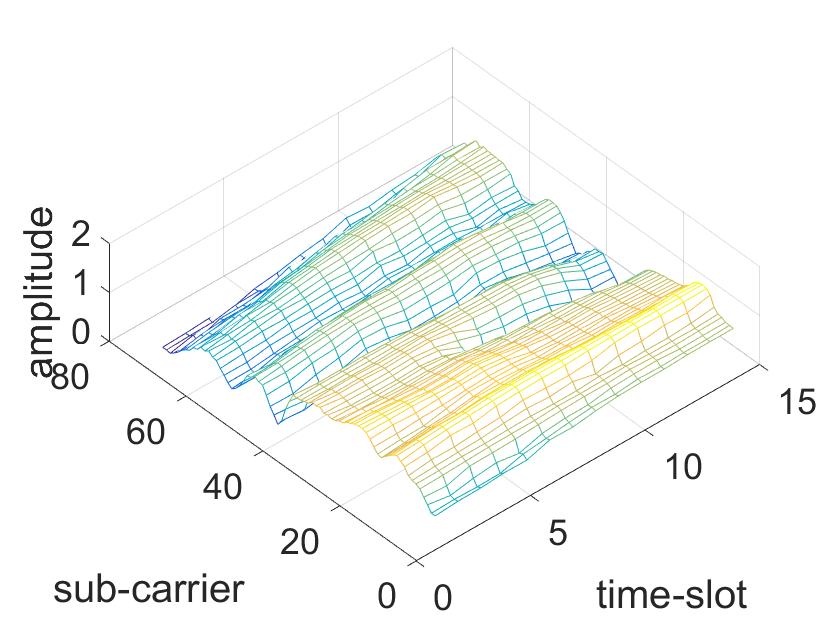}  
			\caption{}
			\label{Speed_G:sub-fourth}
		\end{subfigure}
		\caption{A random resulting sample of the Speed Adaptation Network. (\subref{Speed_G:sub-first}) The input ETU channel with User Speed of 50 km/h. (\subref{Speed_G:sub-second}) The adapted ETU channel to the speed of 25 km/h. (\subref{Speed_G:sub-third}) The adapted ETU channel to the speed of 75 km/h. (\subref{Speed_G:sub-fourth}) The adapted ETU channel to the speed of 100 km/h. }
		\label{Speed_G}
	\end{adjustbox}
\end{figure}
\subsection{Speed Adaptation Network}

Having the generated channel samples at the reference speed, the speed adaptation network will generate an equivalent sample for the desired user speed. The structure of the generator, discriminator, and loss functions are selected as the architecture of StarGAN \cite{choi2018stargan}. 

To train and evaluate the results of the Speed Adaptation Network, we consider the ETU channel model. More specifically, StarGAN is trained considering ETU channel samples with the user speed of $ 50 km/h $ as the reference input, and ETU channel samples with user speeds of $ 25, 75, 100 km/h $ as the target channel images. The desired user speed is fed to the network as a One-hot vector. 

To test the statistical similarity of the resulting 2D channel images, after obtaining the complex time-frequency responses of the channel from the channel images, first, the 2D time-frequency responses are converted to one-dimensional sequences. Since the main effect of the different user speeds is on the time axis, this time we put along all the time slots of the subcarriers for this purpose. (note that for evaluation of the Channel Sample Generator, all sub-carriers of the time-slots are put along).
Then we use the mean autocorrelation and CDM.
The mean autocorrelation of the actual and generated ETU channels for three different user speeds are shown in Figure {\ref{xc_speed}} and Figure {\ref{xc_speed_G}}, respectively. By comparing these two figures, the similarity between the mean autocorrelation function of the generated channels of each speed with its corresponding mean autocorrelation function of the actual channel samples can be verified. This shows the ability of the network in modeling the channels with different user speeds. We have also computed the CDM between the resulting mean autocorrelation functions. The results are listed in Table\ref{tableSpeed}. Again From the table, it is clear that the distance between generated channels and their corresponding measurement is on average an order of magnitude less than the distance between generated channels of one speed and measurements of different speeds. Note that for the speed of $100 km/h$, the CDM shows the best match between the generated and actual samples of  $100 km/h$, however, the value is only 0.4 times the value for the speed of $75 km/h$. This might show that the channel statistics are not very different at high speeds. 

Figure {\ref{Speed_G}} illustrates the results of the Speed Adaptation Network when it got the channel image of 50 km/s as the input and was instructed to generate samples with different user speeds. Figure \ref{Speed_G}(\subref{Speed_G:sub-first}) shows the input channel image for user speed of 50 km/s, Figure \ref{Speed_G}(\subref{Speed_G:sub-second}), is the resulting channel with the User Speed of 25 km/h. As can be seen from the figure, the variations in the time-axis has been reduced. \ref{Speed_G}(\subref{Speed_G:sub-third}), and  \ref{Speed_G}(\subref{Speed_G:sub-fourth}) are the resulting channels with the User Speeds of 75 km/h and 100 km/h, respectively. The increase in the variations of the time-axis in obvious.
\begin{table}[t]
	\centering
	\caption{Cepstral Distance Measure for ETU channels with different user speeds \\}    
	\begin{tabular}{l|l|l|l}
		
		& \makecell{ETU-generated\\25 km/h} & \makecell{ETU-generated\\75 km/h} & \makecell{ETU-generated\\100 km/h} \\
		\hline
		\makecell{\\ ETU-25\\  \\} & $1.56\times 10^{-5}$ & $4.08\times 10^{-4}$ & $8.89\times 10^{-4}$ \\
		\hline
		\makecell{\\ ETU-75\\  \\} & $1.20\times 10^{-4}$ & $2.98\times 10^{-5}$ & $2.28\times 10^{-4}$ \\
		\hline
		\makecell{\\ ETU-100\\  \\} & $4.99\times 10^{-4}$ & $3.79\times 10^{-5}$ & $1.43\times 10^{-5}$
		
	\end{tabular}
	\label{tableSpeed}
\end{table}
\section{Conclusion}
In this paper, a novel propagation channel modeling method based on Deep learning techniques is presented. The time-frequency response of the propagation channel is considered as an image, and the distribution of channel images is modeled using DCGANs. Moreover, the model is extended for measurements having different user speeds. We take advantage of StarGAN as an image-to-image translation technique. The speed adaptation network is trained to learn the effect of different user speeds on channel images. The statistical similarity between the generated and the actual samples are examined with commonly used metrics such as LCR and AFD, as well as a newly introduced metric based on the Cepstral distance between the mean of the autocorrelation functions. 

\bibliography{Refrences}
\bibliographystyle{ieeetr}

\end{document}